\documentclass[runningheads]{llncs}
\usepackage{amsmath}
\usepackage{bm}
\usepackage{amsfonts}
\usepackage{multirow}
\usepackage{colortbl}
\usepackage{enumitem}
\usepackage{tabu}
\usepackage{balance}
\usepackage{footnote}
\makesavenoteenv{tabular}
\makesavenoteenv{table}
\usepackage{tikz}
\usepackage{adjustbox}
\usepackage{svg}
\usepackage{times}
\usepackage{latexsym}
\usepackage{amssymb}
\usepackage{listings}
\usepackage{booktabs}

\newcommand{\suzan}[1]{\textcolor{black}{#1}}
\begin{document}

\setlength{\abovedisplayskip}{1pt}
\setlength{\belowdisplayskip}{1pt}

\title{PARM: A Paragraph Aggregation Retrieval Model for Dense Document-to-Document Retrieval}
\author{Sophia Althammer\inst{1} \and Sebastian Hofst{\"a}tter\inst{1} \and Mete Sertkan\inst{1} \and Suzan Verberne\inst{2} \and Allan Hanbury\inst{1}}
%


\authorrunning{Althammer et al.}
\titlerunning{PARM: A Paragraph Aggregation Retrieval Model}

%
\institute{Institute for Information Systems Engineering, TU Wien, Vienna, Austria
\email{name.surname@tuwien.ac.at}\\
\and
Leiden University, Leiden, Netherlands
\email{s.verberne@liacs.leidenuniv.nl}}

\maketitle              
\begin{abstract}
Dense passage retrieval (DPR) models shoaw great effectiveness gains in first stage retrieval for \suzan{the web domain}. However in the web domain we are in a setting with large amounts of training data and a query-to-passage or a query-to-document retrieval task.
We investigate in this paper dense document-to-document retrieval with limited labelled target data for training, \suzan{in particular} legal case retrieval. In order to use DPR models for document-to-document retrieval, we propose a Paragraph Aggregation Retrieval Model (PARM) which liberates DPR models from their limited input length. PARM retrieves documents on the paragraph-level: for each query paragraph, relevant documents are retrieved based on their paragraphs. Then the relevant results per query paragraph are aggregated into one ranked list for the whole query document. For the aggregation we propose vector-based aggregation with reciprocal rank fusion (VRRF) weighting, which combines the advantages of rank-based aggregation and topical aggregation based on the dense embeddings. Experimental results show that VRRF outperforms rank-based aggregation strategies for dense document-to-document retrieval with PARM. We compare PARM to document-level retrieval and demonstrate higher retrieval effectiveness of PARM for lexical and dense first-stage retrieval on \suzan{two} different legal case retrieval collections. We investigate how to train the dense retrieval model for PARM on limited target data with labels on the paragraph or the document-level. \suzan{In addition,} we analyze the differences of the retrieved results of lexical and dense retrieval with PARM.
\end{abstract}

\section{Introduction}

Dense passage retrieval (DPR) models brought substantial effectiveness gains to information retrieval (IR) tasks in the web domain \cite{Gao2020,karpukhin-etal-2020-dense,xiong2021approximate}. The promise of DPR models is to boost the recall of first stage retrieval by leveraging the semantic information for retrieval as opposed to traditional retrieval models \cite{bm25}, which rely on lexical matching. 
The web domain is a setting with query-to-passage or query-to-document retrieval tasks and a large amount of training data\suzan{, while training data is much more limited in other domains}. Furthermore we see recent advances in neural retrieval remain neglected for document-to-document retrieval despite the task's importance in several, mainly professional, domains \cite{Locke2017automatic,florinapatent,colieesummary,risch2020patentmatch}.

\suzan{In this paper} we investigate \suzan{the effectiveness of dense retrieval models for} document-to-document tasks\suzan{, in particular} legal case retrieval. We focus on first stage retrieval with dense models and therefore aim for a high recall.
The first challenge for DPR models in document-to-document retrieval tasks is the input length of the query documents and of the documents in the corpus. In legal case retrieval the cases tend to be long documents \cite{legalrelevance} with an average length of $1269$ words in the COLIEE case law corpus \cite{colieesummary}. However the input length of DPR models is limited to $512$ tokens \cite{karpukhin-etal-2020-dense} and theoretically bound of how much information of a long text can be compressed into a single vector \cite{luan2020sparse}.
Furthermore we reason in accordance with the literature \cite{passagelevelfordocretrieval,bertpli,passagelevelfordoc,passagelevelinfluence} that relevance between two documents is not only determined by the complete text of the documents, but that a candidate document can be relevant to a query document based on one paragraph that is relevant to one paragraph of the query document.
In the web domain DPR models are trained on up to $500k$ training samples \cite{msmarco16}, whereas in most domain-specific \suzan{collections} 
only a limited amount of hundreds of labelled samples is available \cite{fireaila2,treclegal,colieesummary}.

In this paper we address these challenges by proposing a \textbf{paragraph aggregation retrieval model (PARM)} for dense document-to-document retrieval.
PARM liberates dense passage retrieval models from their limited input length without increasing the computational cost. Furthermore PARM gives insight on which paragraphs the document-level relevance is based, which is beneficial for understanding and explaining the retrieved results.
With PARM the documents are retrieved on the paragraph-level: the query document and the documents in the corpus are split up into their paragraphs and for each query paragraph a ranked list of relevant documents based on their paragraphs is retrieved. The ranked lists of documents per query paragraph need to be aggregated into one ranked list for the whole query document. As PARM provides the dense vectors of each paragraph, we propose \textbf{vector-based aggregation with reciprocal rank fusion weighting (VRRF)} for PARM. VRRF combines the merits of rank-based aggregation \cite{rrf,garciasecodeherrera2014comparing} with semantic aggregation with dense embeddings. We investigate:

\textbf{RQ1} \emph{How does VRRF compare to other aggregation strategies within PARM?}

We find that our proposed aggregation strategy of VRRF for PARM leads to the highest retrieval effectiveness in terms of recall compared to rank-based \cite{rrf,Shaw94combinationof} and vector-based aggregation baselines \cite{li2020parade}.
Furthermore we investigate:

\textbf{RQ2} \emph{How effective is PARM with VRRF for document-to-document retrieval?}

We compare PARM with VRRF to document-level retrieval for lexical and dense retrieval methods on two different test collections for the document-to-document task of legal case retrieval.
We demonstrate that PARM consistently improves the first stage retrieval recall for dense document-to-document retrieval.
Furthermore, dense document-to-document retrieval with PARM and VRRF aggregation outperforms lexical retrieval methods in terms of recall at higher cut-off values.



The success of DPR relies on the size of labelled training data. As we have a limited amount of labelled data as well as paragraph and document-level labels we investigate:


\textbf{RQ3} \emph{How can we train dense passage retrieval models for PARM for document-to-document retrieval most effectively?}

For training DPR for PARM we compare training with relevance labels on the paragraph or document-level. We find that despite the larger size of document-level labelled datasets, the additional training data is not always beneficial compared to training DPR on smaller, but more accurate paragraph-level samples.
%
%
%
%
Our contributions are:
\vspace{-0.2cm}
\begin{itemize}
    \item We propose a \textbf{paragraph aggregation retrieval model (PARM)} for dense document-to-document retrieval and demonstrate higher retrieval effectiveness for dense retrieval with PARM compared to retrieval without PARM and to lexical retrieval with PARM.
    \item We propose \textbf{vector-based aggregation with reciprocal rank fusion weighting (VRRF)} for dense retrieval with PARM and find that VRRF leads to the highest recall for PARM compared to other aggregation strategies.
    \item We investigate training DPR for PARM and compare the impact of fewer, more accurate paragraph-level labels to more, potentially noisy document-level labels.
    \item We publish the code at \url{https://github.com/sophiaalthammer/parm}
\end{itemize}

\section{Related work}


\emph{Dense passage retrieval.} Improving the first stage retrieval with DPR models is a rapidly growing area in neural \suzan{IR}, mostly focusing on \suzan{the web domain}.
Karpukhin et al. \cite{karpukhin-etal-2020-dense} propose dense passage retrieval for open-domain QA using BERT models as bi-encoder for the query and the passage. With ANCE, \suzan{Xiong et al.} \cite{xiong2021approximate} train a DPR model for open-domain QA with sampling negatives from the continuously updated index. Efficiently training DPR models with distillation \cite{hofstatter2021improving} and balanced topic aware sampling \cite{hofstatter2021efficientlytas} has demonstrated to improve the retrieval effectiveness. 
As opposed to this prior work, we move from dense passage to dense document-to-document retrieval and propose PARM to use dense retrieval for document-to-document tasks.

\emph{Document retrieval.} 
The passage level influence for retrieval of documents has been analyzed in multiple works \cite{passagelevelfordocretrieval,passageretrieavllm,passagelevelfordoc,passagelevelinfluence} and shown to be beneficial, but in these works the focus lies on passage-to-document retrieval. Cohan et al. \cite{cohan-etal-2020-specter} present document-level representation learning strategies for ranking, however the input length remains bounded by $512$ tokens and only title and abstract of the document are considered. Abolghasemi et al. \cite{aminecir2022coliee} present multi-task learning for document-to-document retrieval. Liu et al. \cite{smith} propose similar document matching for documents up to a length of $2048$ however here the input length is still bounded and the computational cost of training and using the model is increased.
Different to this prior work, the input length of PARM is not bounded without increasing the computational complexity of the retrieval.

\emph{Aggregation strategies.} Aggregating results from different ranked lists has a long history in IR. Shaw et al. \cite{combmnz,Shaw94combinationof} investigate the combination of multiple result lists by summing the scores. Different rank aggregation strategies like Condorcet \cite{condorcet} or Borda count \cite{borda} are proposed, however it is demonstrated \cite{rrf,comparingscoreaggregation} that reciprocal rank fusion outperforms them.
Ai et al. \cite{neuralpassagemodel} propose a neural passage model for scoring passages for a passage-to-document retrieval task.
Multiple works \cite{akkalyoncu-yilmaz-etal-2019-applying,akkalyoncu-yilmaz-etal-2019-crossdomain,daicallmsmarcodoc,Zhang2021ComparingSA} propose score aggregation for re-ranking with BERT on a passage-to-document task ranging from taking the first passage of a document to the passage of the document with the highest score. 
Different to rank/score-based aggregation approaches, Li et al. \cite{li2020parade} propose vector-based aggregation for re-ranking for a passage-to-document task. Different to our approach they concatenate query and passage and learn a representation for binary classification of the relevance score.
The focus of score/rank aggregation is mainly on federated search or passage-to-document tasks, however we focus on document-to-document retrieval. We have not seen a generalization of aggregation strategies for the query and candidate paragraphs for document-to-document retrieval yet. Different to previous work, we propose to combine rank and vector-based aggregation methods for aggregating the representation of query and candidate documents independently.

\section{Paragraph aggregation retrieval model (PARM)}
\label{chap:methodparagraphlevelret}

In this section we propose PARM as well as the aggregation strategy VRRF for PARM for dense document-to-document retrieval and training strategies.


\subsection{Workflow}

\begin{figure*}[b]
    \centering
    \setlength{\belowcaptionskip}{-20pt} 
    \includegraphics[width=124mm,scale=1.0]{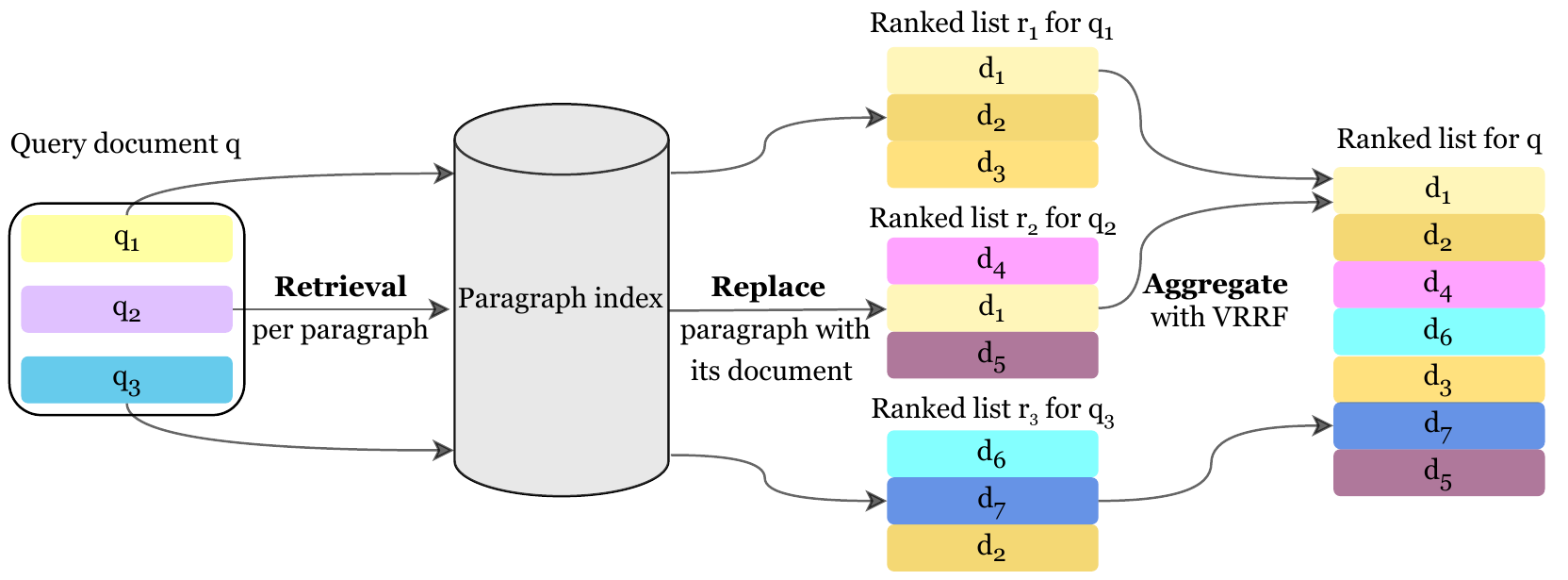}
        \caption{PARM workflow for query document $q$ and retrieved documents $d_1,..,d_7$}
        \label{fig:paragraphlevelretrieval}
\end{figure*}

We use the DPR model \cite{karpukhin-etal-2020-dense} based on BERT \cite{bert} bi-encoders, \suzan{of which} one encodes the query passage $q$, the other one the candidate passage $p$. 
After storing the encoded candidate passages $\hat{p}$ in the index, the relevance score between a query $q$ and a candidate passage $p$ is computed by the dot-product between the encoded query passage $\hat{q}$ and $\hat{p}$.

As the input length of BERT \cite{bert} is limited to $512$ tokens, the input length for the query and the candidate passage for DPR \cite{karpukhin-etal-2020-dense} is also limited by that. The length of query and candidate documents for document-to-document tasks exceeds this input length. For example the average length of a document is $1296$ words for the legal case retrieval collection COLIEE \cite{colieesummary}. We reason that for document-to-document tasks a single paragraph or multiple paragraphs can be decisive for the relevance of a document to another one \cite{passagelevelfordocretrieval,passageretrieavllm,passagelevelfordoc,passagelevelinfluence} and that different paragraphs contain different topics of a document. Therefore we propose a \textbf{paragraph aggregation retrieval model (PARM)}, in order to use DPR models for dense document-to-document retrieval. PARM retrieves relevant documents based on the paragraph-level relevance.

The workflow of PARM is visualized in Fig. \ref{fig:paragraphlevelretrieval}. For the documents in the corpus we 
split each document $d$ into paragraphs $p_1, .., p_{m_d}$ with $m_d$ the number of paragraphs of document $d$. We take the paragraphs of the document as passages for DPR. We index each paragraph $p_j$, $j \in 1, .., m_d$ of each document $d$ in the corpus and attain a paragraph-level index containing the encoded paragraphs $\hat{p}_j$ for all documents $d$ in the corpus. At query time, the query document $q$ is also split up into paragraphs $q_1, ..., q_{n_q}$, where $n_q$ is the number of paragraphs of $q$. For each query paragraph $q_i$ with $i \in 1, .., n_q$ the top $N$ most relevant paragraphs are retrieved from the paragraph-level corpus. 
The result is a ranked list $r_i$ with $i \in 1, .., n_q$ per query paragraph $q_i$ with $N$ relevant paragraphs.  The \suzan{paragraphs} in the ranked lists $r_i$ with $i \in 1, .., n_q$ are replaced by the documents that contain the paragraphs. 
Therefore it is possible that one document occurs multiple times in the list. In order to attain one ranked list for the whole query document $q$, the ranked \suzan{paragraph} lists of retrieved documents $r_1, ..., r_{n_q}$ of each query paragraph $q_i$ with $i \in 1, .., n_q$ need to be aggregated to one ranked list. 

\subsection{Vector-based aggregation with reciprocal rank fusion weighting (VRRF)}
\label{chap:methodaggregation}

%
Multiple works have demonstrated the benefit of reciprocal rank fusion \cite{rrf,garciasecodeherrera2014comparing,comparerrf} for rank-based aggregation of multiple ranked retrieved lists. Using dense retrieval with PARM we have more information than the ranks and scores of the retrieved paragraphs: we have dense embeddings, which encode the semantic meaning of the paragraphs, for each query paragraph and the retrieved paragraphs. In order to make use of this additional information for aggregation, we propose \textbf{vector-based aggregation with reciprocal rank fusion weighting (VRRF)}, which extends reciprocal rank fusion for neural retrieval. VRRF combines the advantages of reciprocal rank fusion with relevance signals of semantic aggregation using the dense vector embeddings.

In \textbf{VRRF} we aggregate documents using the dense embeddings $\hat{p_i}$ of the passages $p_i$, which are from the same document $d$ and which are in the retrieved list $r_i$ with $i \in 1, .., n_q$, with a weighted sum, taking the reciprocal rank fusion score \cite{rrf} as weight. The dense embeddings $\hat{q_i}$ of each query paragraph $q_i$ with $i \in 1, .., n_q$ are aggregated by adding the embeddings without a weighting:
\begin{align*}
    \hat{q} = \sum_{i=1}^{n_q} \hat{q_i} && \hat{d} = \sum_{i=1}^{n_q} \sum_{p \in d, d \in r_i} rrf(q_i, p_i) \ \hat{p_i}
\end{align*}

We compute the relevance score between query and candidate document with the dot-product between the aggregated embedding of query $\hat{q}$ and candidate document $\hat{d}$.

To confirm the viability of VRFF aggregation, we propose simple baselines: \textbf{VRanks} and \textbf{VScores}, where the paragraph embeddings $\hat{p_i}$ of $d$ are aggregated with the rank or the score of the passage $p_i$ as weight.

\subsection{Training strategies}
\label{chap:methodparalevelvsdoc}

As we have a limited amount of labelled target data, we examine how to effectively train a DPR model for PARM with the training collections at hand. We assume that we have test collections consisting of documents with clearly identifiable paragraphs, with relevance assessments on either the paragraph or the document-level.\newline
\emph{Paragraph-level training.}
For the paragraph-level labelled training we take the relevant paragraphs in the training set as positives and sample random negatives from the paragraphs in the corpus. Here we sample as many negatives as we have positive samples for each query paragraph, thereby balancing the training data.\newline
\emph{Document-level training.}
For the document-level labelled training the collection contains query documents and a corpus of documents with relevance assessments for each query document. We sample negative documents randomly from the corpus.
In order to use the document-level labelled collection for training the DPR model, we split up the query document as well as the positive documents into its paragraphs and consider each paragraph of the query document relevant to each paragraph of each positive document. Equivalently we consider each paragraph of a negative document irrelevant to each query paragraph. 
As on average each document in the COLIEE dataset \cite{colieesummary} contains $42.44$ paragraphs, one relevant document leads to $42 \cdot 20=840$ paragraph-level labels containing one positive and one negative sample to a query paragraph. Therefore this method greatly increases the number of paragraph-level annotations, however this comes with the risk of potentially noisy labels \suzan{\cite{akkalyoncu2019}}.

\section{Experiment Design}


\subsection{Training and test collections}
\label{sec:data}

We focus on the document-to-document task of legal case retrieval because of the importance for the legal domain \cite{Locke2018caselaw,Locke2017automatic,thuircoliee,bertpli} which facilitates the availability of training collections with relevance annotations on the paragraph and the document-level \cite{colieesummary}. For training the DPR models, we introduce paragraph and document-level labelled collections. For the evaluation we use the document-level collections.\newline
\emph{Paragraph-level labelled collections.} COLIEE \cite{colieesummary} is a competition for legal information extraction and retrieval which provides datasets for legal case retrieval and case entailment.
\suzan{Task 2 of} COLIEE 2020 \cite{colieesummary} \suzan{provides} a training and test collection for legal case entailment. It contains relevance labels on the legal case paragraph level, given a query claim, a set of 
\suzan{candidate claims} to the query claim as well as relevance labels for the candidate claims. We denote these sets with \emph{COLIEEPara train/test}.\newline
\emph{Document-level labelled collections.} \suzan{In Task 1 of} COLIEE 2021 \cite{colieesummary}, the legal case retrieval task, query cases with their relevance judgements on the document-level are provided together with a corpus of candidate documents.
We divide the training set of COLIEEDoc into a training and validation set. The validation set contains the last $100$ queries of the training set from query case $550$ to $650$. We will denote the training, validation and test collection with \emph{COLIEEDoc train/val/test}.
For a broader evaluation, we evaluate our models additionally on the CaseLaw collection \cite{Locke2017automatic}. It contains a corpus of legal cases, query cases and their relevance judgements for legal case retrieval.
~\\
\emph{Data pre-processing.} For COLIEEDoc, we remove the French versions of the cases, we divide the cases into an introductory part, a summary, if it contains one, and its claims, which are indicated by their numbering. As indicated in Table \ref{tab:datasetstat}, the paragraphs have an average length of $84$ words and $96.2\%$ of the paragraphs are not longer than $512$ words.
The CaseLaw dataset is split along the line breaks of the text and merged to paragraphs by concatenating sentences until the paragraphs exceed the length of $200$ words.

\subsection{Baselines}

As baseline we use the lexical retrieval model BM25 \cite{bm25}.
For BM25 we use ElasticSearch\footnote{\url{https://github.com/elastic/elasticsearch}} with parameters $k=1.3$ and $b=0.8$, which we optimized on COLIEEDocval.
\begin{table}[t!]
    \centering
    \caption{Statistics of paragraph- and document-level labelled collections.}
    \label{tab:datasetstat}
    \setlength\tabcolsep{2pt}
    \begin{tabular}{lllcccccccc}
       \toprule
       \multirow{1}{*}{\textbf{Labels}}   &
       \multirow{1}{*}{\textbf{Dataset}}   & \multirow{1}{*}{\textbf{Train/}}
         & \multicolumn{6}{c}{\textbf{Statistics}}\\
       &&\textbf{Test}& \small{\# queries}  & \small{$\varnothing$ \# docs} & \small{$\varnothing$ \# rel} & \small{$\varnothing$ para} &\% para $<$& \small{$\varnothing$ \# para} \\
       &&&&& \small{docs} &\small{length} &$512$ words& \\
        \midrule
        \arrayrulecolor{lightgray}
        \multirow{2}{*}{Para} &
          \multirow{1}{*}{COLIEEPara} & Train&325&32.12&1.12 &102&95.5\%&-\\ 
            &\multirow{1}{*}{COLIEEPara}& Test&100&32.19&1.02&117&95.2\%&-\\
         \midrule
        \arrayrulecolor{black}
        \multirow{3}{*}{Doc} &
          \multirow{1}{*}{COLIEEDoc} & Train&650&4415&5.17&84&96.2\%&44.6 \\ 
            &\multirow{1}{*}{COLIEEDoc}& Test&250&4415&3.60&92&97.8\%&47.5\\
        &\multirow{1}{*}{CaseLaw} & Test &100&63431&7.2&219&91.3\%&7.5 \\ 
        \arrayrulecolor{black}
        \bottomrule
    \end{tabular}
    \vspace{-0.5cm}
\end{table}
\emph{VRRF aggregation for PARM (RQ1).}
In order to investigate the retrieval effectiveness of our proposed aggregation strategy VRFF for PARM, we compare VRRF to the commonly used score-based aggregation strategy CombSum \cite{Shaw94combinationof} and rank-based aggregation strategy of reciprocal rank fusion (RRF) \cite{rrf} for PARM.
As baselines for vector-based aggregation, we investigate VSum, VMin, VMax, VAvg, which are originally proposed by Li et al. \cite{li2020parade} for re-ranking on a passage-to-document retrieval task.
In order to use VSum, VMin, VMax, VAvg in the context of PARM, we aggregate independently the embeddings of both, the query and the candidate document. In contrast to Li et al. \cite{li2020parade} we aggregate the query and paragraph embeddings independently and score the relevance between aggregated query and aggregated candidate embedding after aggregation. The learned aggregation methods of CNN and Transformer proposed by Liu et al. \cite{li2020parade} are therefore not applicable to PARM, as they learn a classification on the embedding of the concatenated query and paragraph.\newline
%
\emph{PARM VRRF for dense document-to-document retrieval (RQ2).}
In order to investigate the retrieval effectiveness of PARM with VRRF for dense document-to-document retrieval, we compare PARM to document-level retrieval on two document-level collections (COLIEEDoc and CaseLaw). Because of the limited input length, the document-level retrieval either reduces to retrieval based on the First Passage (FirstP) or the passage of the document with the maximum score (MaxP) \cite{akkalyoncu-yilmaz-etal-2019-applying,Zhang2021ComparingSA}. In order to separate the impact of PARM for lexical and dense retrieval methods, we also use PARM with BM25 as baseline. For PARM with BM25 we also investigate which aggregation strategy leads to the highest retrieval effectiveness in order to have a strong baseline. As BM25 does not provide dense embeddings only rank-based aggregation strategies are applicable.\newline
%
%
\emph{Paragraph and document-level labelled training (RQ3).}
We train a DPR model on a paragraph- and another document-level labelled collection and compare the retrieval performance of PARM for document-to-document retrieval. As bi-encoders for DPR we choose BERT \cite{bert} and LegalBERT \cite{chalkidis-etal-2020-legalbert}.
We train DPR on the paragraph-level labelled collection COLIEEPara train and additionally on the document-level labelled collection COLIEEDoc train as described in Section \ref{chap:methodparalevelvsdoc}. We use the public code\footnote{\url{https://github.com/facebookresearch/DPR}} and train DPR according to Karpukhin et al. \cite{karpukhin-etal-2020-dense}.
We sample the negative paragraphs randomly from randomly sampled negative documents and take the $20$ paragraphs of a positive document as positive samples, which have the highest BM25 score to the query paragraph. This training procedure lead to the highest recall compared to training with all positive paragraphs or with BM25 sampled negative paragraphs.
We also experimented with the DPR model pre-trained on open-domain QA as well as TAS-balanced DPR model \cite{hofstatter2021efficientlytas}, but initial experiments did not show a performance improvement.
We train each DPR model for $40$ epochs and take the best checkpoint according to COLIEEPara test/COLIEEDoc val. We use batch size of $22$ and a learning rate of $2*10^{-5}$, after comparing three commonly used learning rates ($2*10^{-5}$, $1*10^{-5}$, $5*10^{-6}$) for  \cite{karpukhin-etal-2020-dense}.





\section{Results and Analysis}
\label{chap:resultsandanalysis}

We evaluate the first stage retrieval performance with nDCG@10, recall@100, recall@500 and recall@1k using pytrec\textunderscore eval. 
We focus our evaluation on recall 
because the recall performance of the first stage retrieval bounds the ranking performance after re-ranking the results \suzan{in the second stage} for a higher precision. We do not compare our results to \suzan{the reported state-of-the-art results} as they rely on re-ranked results and do not \suzan{report} evaluation results after the first stage retrieval.

\subsection{RQ1: VRRF aggregation for PARM}
\label{sec:resultsaggregation}

As we propose vector-based aggregation with reciprocal rank fusion weighting (VRRF) for PARM, we first investigate:\newline (\textbf{RQ1}) \emph{How does VRRF compare to other aggregation strategies within PARM?}
~\\
We compare VRRF, which combines dense-vector-based aggregation with rank-based weighting, to score/rank-based and vector-based aggregation methods for PARM. The results in Table \ref{tab:aggregation} show that VRRF outperforms all rank and vector-based aggregation approaches for the dense retrieval results of DPR PARM with BERT and LegalBERT. For the lexical retrieval BM25 with PARM, only rank-based aggregation approaches are feasible, here RRF shows the best performance, which will be our baseline for RQ2.

\vspace{-0.5cm}

\begin{table*}
    \small
    \centering
    \caption{Aggregation comparison for PARM on COLIEEval, VRRF shows best results for dense retrieval, stat. sig. difference to RRF w/ paired t-test (p$<$0.05) denoted with $\dagger$, Bonferroni correction with n$=$7. For BM25 only rank-based methods applicable.}
    \label{tab:aggregation}
    \begin{tabular}{l!{\color{lightgray}\vrule}ccc!{\color{lightgray}\vrule}ccc!{\color{lightgray}\vrule}ccc}
       \toprule
       \multirow{2}{*}{\textbf{Aggregation}}   &
       \multicolumn{3}{c}{\textbf{BM25}}&
       \multicolumn{3}{c}{\textbf{DPR BERT}}&
       \multicolumn{3}{c}{\textbf{DPR LegalBERT}}\\
       & \small{R@100} & \small{R@500} & \small{R@1K} & \small{R@100} & \small{R@500} & \small{R@1K} & \small{R@100} & \small{R@500} & \small{R@1K} \\
        \midrule
        \multicolumn{5}{l}{\textbf{Rank-based}} \\
        CombSum \cite{Shaw94combinationof}  &.5236&.7854&.8695 &.4460&.7642&.8594 & .5176&	.7975&	.8882\\
        RRF \cite{rrf} &\textbf{.5796}&\textbf{.8234}&	\textbf{.8963}&.5011&.8029&.8804& \textbf{.5830}&.8373&.9049  \\
        \arrayrulecolor{lightgray}
        \midrule
         \multicolumn{5}{l}{\textbf{Vector-based}} \\
        VAvg \cite{li2020parade}&-&-&-&$.1908^{\dagger}$&$.4668^{\dagger}$&$.6419^{\dagger}$&$.2864^{\dagger}$&$.4009^{\dagger}$&$.7466^{\dagger}$  \\
        VMax \cite{li2020parade}&-&-&-&$.3675^{\dagger}$&$.6992^{\dagger}$&$.8273^{\dagger}$&$.4071^{\dagger}$&$.6587^{\dagger}$&$.8418^{\dagger}$  \\
        VMin \cite{li2020parade}&-&-&-&$.3868^{\dagger}$&$.6869^{\dagger}$&$.8295^{\dagger}$&$.4154^{\dagger}$&$.6423^{\dagger}$&$.8465^{\dagger}$ \\
        VSum \cite{li2020parade}&-&-&-&.4807&$.7496^{\dagger}$&.8742&$.5182^{\dagger}$&$.8069$&$.8882$ \\
        \arrayrulecolor{lightgray}
        \midrule
        \multicolumn{10}{l}{\textbf{Vector-based with rank-based weights (Ours)}} \\
        VScores &-&-&-&.4841&$.7616^{\dagger}$&.8709&$.5195^{\dagger}$&$.8075^{\dagger}$&$.8882^{\dagger}$  \\
        VRanks &-&-&-&.4826&$.7700^{\dagger}$&$.8804$&$.5691^{\dagger}$&$.8212$&$.8980$ \\
        VRRF &-&-&-&\textbf{.5035}&$\textbf{.8062}^{\dagger}$&$\textbf{.8806}$&$\textbf{.5830}^{\dagger}$&$\textbf{.8386}^{\dagger}$&$\textbf{.9091}^{\dagger}$ \\
        \arrayrulecolor{black}
        \bottomrule
    \end{tabular}
    \vspace{-0.5cm}
\end{table*}



\begin{table*}[t]
    \centering
    \caption{Doc-to-doc retrieval results for PARM and Document-level retrieval. No comparison to \suzan{results reported in prior work} as those rely on re-ranking, \suzan{while we evaluate} only first stage retrieval evaluation. \textit{nDCG cutoff at 10, stat. sig. difference to BM25 Doc w/ paired t-test (p $<$ 0.05) denoted with $\dagger$ and Bonferroni correction with n$=$12, effect size $>$0.2 denoted with $\ddagger$.}}
    \label{tab:rq23}
   \begin{tabular}{ll!{\color{lightgray}\vrule}llll!{\color{lightgray}\vrule}llll}
       \toprule
        \multicolumn{1}{l}{\textbf{Model}}
       & \textbf{Retrieval}& 
       \multicolumn{4}{c}{\textbf{COLIEEDoc test}} &\multicolumn{4}{c}{\textbf{CaseLaw}}\\
       && \small{nDCG} & \small{R@100} & \small{R@500} & \small{R@1K} & \small{nDCG}&\small{R@100}  & \small{R@500} & \small{R@1K}  \\
        \midrule
        \multicolumn{6}{l}{\textbf{BM25}} \\
         \multirow{2}{*}{BM25}     & Doc & \textbf{.2435} &.6231&.7815&.8426
        &\textbf{.2653} &.4218&.5058&.5438\\
               & PARM RRF& $.1641^{\dagger\ddagger}$&$\textbf{.6497}^{\dagger\ddagger}$&$.8409^{\dagger\ddagger}$&$.8944^{\dagger\ddagger}$&$.0588^{\dagger\ddagger}$&$.3362^{\dagger\ddagger}$&$.5716^{\dagger\ddagger}$&$.6378^{\dagger\ddagger}$\\
        \arrayrulecolor{black}
        \midrule
        \arrayrulecolor{lightgray}
        \multicolumn{9}{l}{\textbf{DPR}} \\
          \multirow{4}{*}{BERT para} & Doc FirstP & $.0427^{\dagger\ddagger}$&$.3000^{\dagger\ddagger}$&$.5371^{\dagger\ddagger}$	&$.6598^{\dagger\ddagger}$ &$.0287^{\dagger\ddagger}$&$.0871^{\dagger\ddagger}$&$.1658^{\dagger\ddagger}$&$.2300^{\dagger\ddagger}$\\
          & Doc MaxP& $.0134^{\dagger\ddagger}$&$.1246^{\dagger\ddagger}$&$.5134^{\dagger\ddagger}$&$.6201^{\dagger\ddagger}$&$.0000^{\dagger\ddagger}$&$.0050^{\dagger\ddagger}$&$.4813^{\dagger\ddagger}$&$.4832^{\dagger\ddagger}$\\
                        & PARM RRF &$.0934^{\dagger\ddagger}$& $.5765^{\dagger\ddagger}$&$.8153^{\dagger\ddagger}$&$.8897^{\dagger\ddagger}$&$.0046^{\dagger\ddagger}$&$.1720^{\dagger\ddagger}$&$.5019^{\dagger\ddagger}$&$.5563^{\dagger}$\\
                        & PARM VRRF &$.0952^{\dagger\ddagger}$& $.5786^{\dagger\ddagger}$&$.8132^{\dagger\ddagger}$&$.8909^{\dagger\ddagger}$&$.1754^{\dagger\ddagger}$&$.3855^{\dagger\ddagger}$&$.5328^{\dagger\ddagger}$&$.5742^{\dagger\ddagger}$\\
        \midrule
        \arrayrulecolor{lightgray}
          \multirow{4}{*}{LegalBERT para} & Doc FirstP &$.0553^{\dagger\ddagger}$&$.2447^{\dagger\ddagger}$&$.4598^{\dagger\ddagger}$&$.5657^{\dagger\ddagger}$&$.0397^{\dagger\ddagger}$&$.0870^{\dagger\ddagger}$&$.1844^{\dagger\ddagger}$&$.2248^{\dagger\ddagger}$\\
                      & Doc MaxP &$.0073^{\dagger\ddagger}$&$.0737^{\dagger\ddagger}$&$.3970^{\dagger\ddagger}$&$.5670^{\dagger\ddagger}$&$.0000^{\dagger\ddagger}$&$.0050^{\dagger\ddagger}$&$.4846^{\dagger\ddagger}$&$.4858^{\dagger\ddagger}$\\
                       & PARM RRF &$.1280^{\dagger\ddagger}$ &$.6370$&$.8308^{\dagger\ddagger}$&$.8997^{\dagger\ddagger}$&$.0177^{\dagger\ddagger}$&$.2595^{\dagger\ddagger}$&$.5446^{\dagger\ddagger}$&$.6040^{\dagger\ddagger}$\\
                       & PARM VRRF &$.1280^{\dagger\ddagger}$ &$.6396$&$.8310^{\dagger\ddagger}$&$.9023^{\dagger\ddagger}$&$.0113^{\dagger\ddagger}$&$\textbf{.4986}^{\dagger\ddagger}$&$.5736^{\dagger\ddagger}$&$.6340^{\dagger\ddagger}$\\
        \midrule
        \arrayrulecolor{lightgray}
        \multirow{4}{*}{LegalBERT doc} & Doc FirstP &$.0682^{\dagger\ddagger}$&$.3881^{\dagger\ddagger}$&$.6187^{\dagger\ddagger}$&$.7361^{\dagger\ddagger}$&$.0061^{\dagger\ddagger}$&$.0050^{\dagger\ddagger}$&$.4833^{\dagger\ddagger}$&$.4866^{\dagger\ddagger}$\\
                      & Doc MaxP &$.0008^{\dagger\ddagger}$&$.0302^{\dagger\ddagger}$&$.2069^{\dagger\ddagger}$&$.2534^{\dagger\ddagger}$&$.0022^{\dagger\ddagger}$&$.0050^{\dagger\ddagger}$&$.4800^{\dagger\ddagger}$&$.4833^{\dagger\ddagger}$\\
                       & PARM RRF &$.1248^{\dagger\ddagger}$ &$.6086^{\dagger}$&$.8394^{\dagger\ddagger}$&$.9114^{\dagger\ddagger}$&$.0117^{\dagger\ddagger}$&$.2277^{\dagger\ddagger}$&$.5637^{\dagger\ddagger}$&$.6265^{\dagger\ddagger}$\\
                       & PARM VRRF &$.1256^{\dagger\ddagger}$ &$.6127^{\dagger}$&$\textbf{.8426}^{\dagger\ddagger}$&$\textbf{.9128}^{\dagger\ddagger}$&$.2284^{\dagger\ddagger}$&$.4620^{\dagger\ddagger}$&$\textbf{.5847}^{\dagger\ddagger}$&$\textbf{.6402}^{\dagger\ddagger}$\\
         \arrayrulecolor{black}
        \bottomrule
    \end{tabular}
\end{table*}

\newpage

\subsection{RQ2: PARM VRRF vs Document-level retrieval}

As we propose PARM VRRF for document-to-document retrieval, we investigate:\newline
(\textbf{RQ2}) \emph{How effective is PARM with VRRF for document-to-document retrieval?}
~\\
We evaluate and compare PARM and document-level retrieval for lexical and dense retrieval methods on the two test collections (COLIEEDoc and CaseLaw) for document-to-document retrieval in Table \ref{tab:rq23}.
%
For BM25 we find that PARM-based retrieval outperforms document-level retrieval at each recall stage, except for R@100 on CaseLaw.

For dense retrieval we evaluate DPR models with BERT trained solely on the paragraph-level labels and with LegalBERT trained on the paragraph-level labels (denoted with LegalBERT para) and with additional training on the document-level labels (denoted with LegalBERT doc). For dense document-to-document retrieval PARM consistently outperforms document-level retrieval for all performance metrics for both test collections. Furthermore PARM aggregation with VRRF outperforms PARM RRF in nearly all cases. Overall we find that LegalBERTdoc-based dense retrieval with PARM VRRF achieves the highest recall at high ranks.
When comparing the nDCG@10 evaluation we find that PARM lowers the nDCG@10 score for BM25 as well as for dense retrieval. Therefore we suggest that PARM is beneficial for first stage retrieval, so that in the re-ranking stage the overall ranking can be improved.

\begin{figure}[t]
\begin{minipage}{0.52\textwidth} 
\centering
    \includegraphics[width=1.0\textwidth]{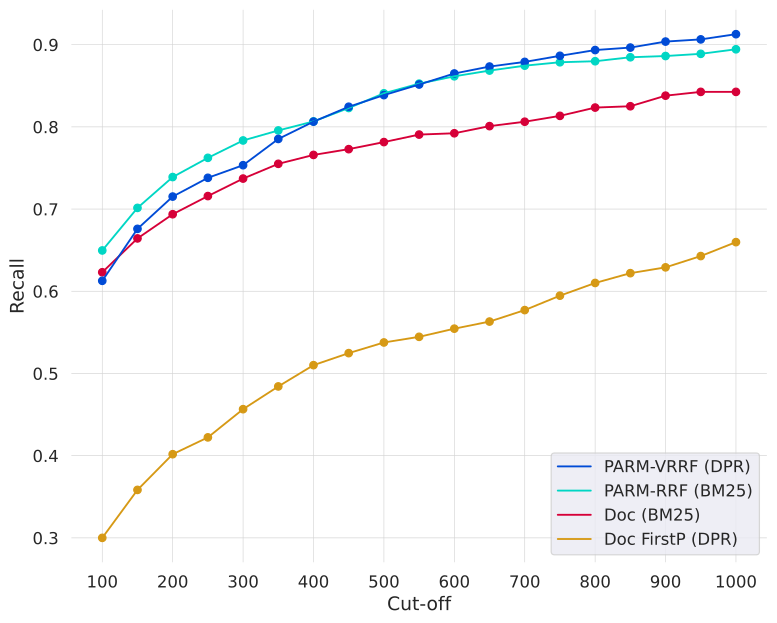}
        \caption{Recall at different cut-off values for PARM-VRRF (DPR) and PARM-RRF (BM25) and Document-level retrieval with BM25 and DPR for COLIEEDoc test.}
        \label{fig:evalparmretrieval}
\end{minipage}%
\hfill
\begin{minipage}{.46\textwidth}
\centering
\label{tab:howmanyrel}
    \begin{tabular}{l!{\color{lightgray}\vrule}cc!{\color{lightgray}\vrule}cc}
       \toprule
       & \multicolumn{2}{c}{\textbf{COLIEEDoc}}&  \multicolumn{2}{c}{\textbf{CaseLaw}}\\
       & \small{BM25} & \small{DPR} & \small{BM25} & \small{DPR} \\
        \midrule
        \multicolumn{1}{l}{\textbf{Total}} \\
         relevant & 900  &900&720&720\\ 
         PARM & 892  &896&578&545\\
         Doc & 751  &662&419&199\\
        \midrule
        \arrayrulecolor{black}
        \multicolumn{1}{l}{\textbf{Sets}} \\
         PARM $\cap$ Doc & 750  &661&417&196\\ 
         PARM\textbackslash Doc &142&235&161&349\\
         Doc\textbackslash PARM &1&1&2&3\\
        \arrayrulecolor{black}
        \bottomrule
    \end{tabular}
    \caption{Number of relevant documents retrieved in comparison between PARM and Doc-level retrieval for COLIEEDoc and CaseLaw with BM25 or LegalBERT\textunderscore doc-based DPR.}
\end{minipage} 
\vspace{-0.5cm}
\end{figure}

In Figure \ref{fig:evalparmretrieval} we \suzan{show} the recall at different cut-off values for PARM-VRRF with DPR (based on LegalBERTdoc) and PARM-RRRF with BM25 compared to document-level retrieval (Doc FirstP) of BM25/DPR. When comparing PARM to document-retrieval, we can see a clear gap between the performance of document-level retrieval and PARM for BM25 and for DPR. Furthermore we see that dense retrieval (PARM-VRRF DPR) outperforms lexical retrieval (PARM-RRF BM25) at cut-off values above $500$.

In order to analyze the differences \suzan{between} PARM and document-level retrieval further, we analyze in Figure 3
, how many relevant documents are retrieved with PARM or with document-level retrieval with lexical (BM25) or dense methods (DPR). Furthermore we investigate how many relevant documents are retrieved by both PARM and document-level retrieval (PARM $\cap$ Doc), and how many relevant documents are retrieved only with PARM and not with document-level retrieval (PARM\textbackslash Doc) and vice versa (Doc\textbackslash PARM).
When comparing the performance of PARM and document-level retrieval, we find that PARM retrieves more relevant documents in total for both test collections.
PARM retrieves $142-380$ of the relevant documents that did not get retrieved with document-level retrieval (PARM\textbackslash Doc), which are $15-52\%$ of the total number of relevant documents. This analysis demonstrates that PARM \suzan{largely} retrieves many of relevant documents that are not retrieved with document-level retrieval.
We conclude that PARM is not only beneficial for dense but also for lexical retrieval. 




\subsection{RQ3: Paragraph-level vs Document-level Labelled Training}
\label{sec:resultsrq1}

As labelled in-domain data for document-to-document retrieval tasks is limited, we ask: (\textbf{RQ3}) \emph{How can we train dense passage retrieval models for PARM for document-to-document retrieval most effectively?} We compare the retrieval performance for BERT-based and LegalBERT-based dense retrieval models in Table \ref{tab:rq1}, which are either trained solely on the paragraph-level labelled collection or additionally trained on the document-level labelled collection. The \suzan{upper part of the table shows} that for BERT the additional training data on document-level improves the retrieval performance for document-level retrieval, but harms the performance for PARM RRF and PARM VRRF. For LegalBERT the additional document-level training data highly improves the performance of document-retrieval. For PARM the recall is improved at higher cut-off values ($@500$, $@1000$) for a cut-off.
Therefore we consider the training on document-level labelled data beneficial for dense retrieval based on LegalBERT.
This reveals that it is not always better to have more, potentially noisy data, for BERT-based dense retrieval the training with fewer, but accurate paragraph-level labels is more beneficial for overall document-to-document retrieval with PARM.





\begin{table}[t!]
    \centering
    \caption{Paragraph- and document-level labelled training of DPR. Document-level labelled training improves performance at high ranks for LegalBERT.}
    \label{tab:rq1}
    \setlength\tabcolsep{1.5pt}
    \begin{tabular}{ll!{\color{lightgray}\vrule}l!{\color{lightgray}\vrule}rrrrr}
       \toprule
        \multirow{2}{*}{\textbf{Model}}
       & 
       \multirow{2}{*}{\textbf{Retrieval}}   &
       \multirow{1}{*}{\textbf{Train}}  &
       \multicolumn{4}{c}{\textbf{COLIEEDoc val}}\\
       &&\textbf{Labels} & \small{R@100}  & \small{R@200} & \small{R@300} & \small{R@500} & \small{R@1K}  \\
        \midrule
        \arrayrulecolor{lightgray}
        \multicolumn{7}{l}{\textbf{DPR Retrieval}} \\
          \multirow{6}{*}{BERT} & Doc FirstP& para& .3000&	.4018&.4566&.5371&.6598 \\ 
          & Doc FirstP& + doc & .3800&.4641&.5160 &.6054	&.7211\\
            & PARM RRF& para & .5765&.6879&.7455&.8153&.8897\\
            & PARM RRF& + doc&.5208&.6502	&.7100	&.7726	&.8660 \\
            & PARM VRRF& para&.5786&.6868&.7505&.8132&.8909 \\
            & PARM VRRF& + doc&.5581&.6696&.7298&.7970&.8768\\
         \midrule
          \multirow{6}{*}{LegalBERT} & Doc FirstP& para& .2447&.3286&.3853	&.4598	&.5657 \\
          & Doc FirstP& + doc& .3881&.4665&.5373&.6187&.7361 \\
            & PARM RRF& para&.6350&.7323&.7834&.8308&.8997\\
            & PARM RRF& + doc&.6086&.7164&.7561&.8394&.9114 \\
            & PARM VRRF& para&\textbf{.6396}&\textbf{.7325}&\textbf{.7864}&.8310&.9023 \\
            & PARM VRRF& + doc&.6098&.7152&.7520&\textbf{.8396}&\textbf{.9128}\\
        \arrayrulecolor{black}
        \bottomrule
    \end{tabular}
\end{table}

\subsection{Analysis of paragraph relations}

\begin{figure}[t]
\begin{minipage}{0.49\textwidth} 
\centering
\includegraphics[width=1.0\textwidth]{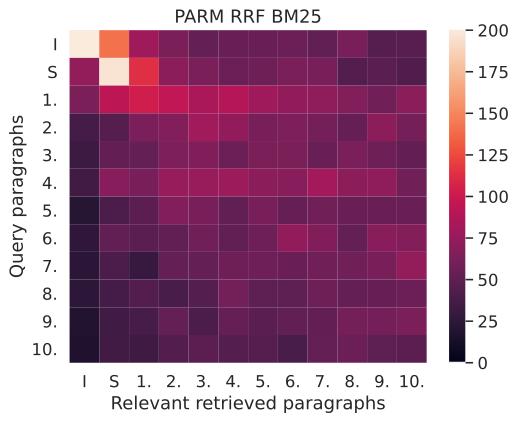}
\end{minipage}%
\begin{minipage}{.49\textwidth}
\centering
\includegraphics[width=1.0\textwidth]{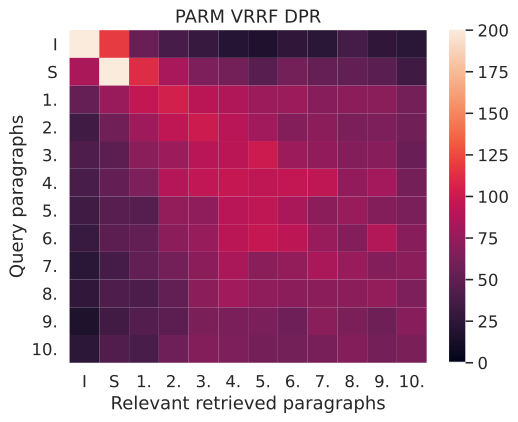}
\end{minipage} 
\caption{Heatmap for PARM retrieval with BM25 or DPR visualizing which query paragraph how often retrieves which paragraph from a relevant document. I denotes the introduction, S the summary, 1.-10. denote the claims 1.-10. of COLIEEDoc test.}
\label{fig:heatmapdpr}
\vspace{-0.5cm}
\end{figure}

With our proposed paragraph aggregation retrieval model for dense document-to-document retrieval we can analyze on which paragraphs the document-level relevance is based.
To gain more insight in what the dense retrieval model learned to retrieve on the paragraph-level with PARM, we analyze which query paragraph retrieves which paragraphs from relevant documents with dense retrieval with PARM and compare it to lexical retrieval with PARM. In Figure \ref{fig:heatmapdpr}, a heatmap visualizes which query paragraph how often retrieves which paragraph from a relevant document with PARM BM25 or PARM DPR on the COLIEEDoc test set.
As introduced in Section \ref{sec:data}, the legal cases in COLIEEDoc contain an introduction, a summary and claims as paragraphs. For the introduction (I) and the summary (S) we see the paragraph relation for lexical and dense retrieval that both methods retrieve also more introductions and summaries from the relevant documents. We reason this is due to the special structure of the introduction and the summary which is distinct to the claims. For the query paragraphs 1.-10. we see that PARM DPR seems to focus on to the diagonal different to PARM BM25. This means for example that the first paragraph retrieves more first paragraphs from relevant documents than they retrieve other paragraphs. As the claim numbers are removed in the data pre-processing, this focus relies on the textual content of the claims. This paragraph relation suggests that there is a topical or hierarchical structure in the claims of legal cases, which is learned by DPR and exhibited with PARM. This structural component can not be exhibited with document-level retrieval.




\section{Conclusion}


In this paper we address the challenges of using dense passage retrieval (DPR) in first stage retrieval for document-to-document tasks with limited labelled data.
We propose the paragraph aggregation retrieval model (PARM), which liberates dense passage retrieval models from their limited input length and which takes the paragraph-level relevance for document retrieval into account. We demonstrate on two test collections higher first stage recall for dense document-to-document retrieval with PARM \suzan{than with document-level retrieval}. We also show that dense retrieval with PARM outperforms lexical retrieval \suzan{with BM25} in terms of recall at higher cut-off values. As part of PARM we propose the novel vector-based aggregation with reciprocal rank fusion weighting (VRFF), which combines the advantages of rank-based aggregation with RRF \cite{rrf} and topical aggregation with dense embeddings. We demonstrate the highest retrieval effectiveness for PARM with VRRF aggregation compared to rank and vector-based aggregation baselines.
Furthermore we investigate how to train dense retrieval models for dense document-to-document retrieval with PARM.
We find the interesting result that training DPR models on more, but noisy document-level data does not always lead to overall higher retrieval performance compared to training on less, but more accurate paragraph-level labelled data.
Finally, we analyze how PARM retrieves relevant paragraphs and find that the dense retrieval model learns a structural paragraph relation which it exhibits with PARM and therefore benefits the retrieval effectiveness.

\section*{Acknowledgements}
This work was supported by EU Horizon 2020 ITN/ETN on Domain Specific Systems for Information Extraction and Retrieval(ID:860721).

%
%
%
\bibliographystyle{splncs04}
\bibliography{mybibliography}
%
%
%
%
%
\end{document}